\documentclass[12pt,longbibliography,nofootinbib]{article}
\usepackage{authblk}
\usepackage[bookmarksnumbered, colorlinks, plainpages]{hyperref}
\usepackage{amsmath, amsthm, amscd, amsfonts, amssymb, graphicx, color, booktabs}

\usepackage{fancyhdr}  
\usepackage{cite}
\usepackage{slashed}
\usepackage[]{units}
\usepackage{bm}

\hypersetup{
    colorlinks=true,       
    linkcolor=blue,          
    citecolor=blue,        
    filecolor=blue,      
    urlcolor=black           
}
\usepackage{cleveref}

\numberwithin{equation}{section}
\definecolor{email}{rgb}{0.00,0.00,0.84}

\fancypagestyle{firstpage}
{
    \fancyhead[L]{}    
    \fancyhead[R]{MIT-CTP/5697}
}

\begin{document}
\setcounter{page}{1}

\newcommand{\Vcb}[0]{\ensuremath{|V_{cb}|}}
\newcommand{\Vub}[0]{\ensuremath{|V_{ub}|}}
\newcommand{\BtoDstar}{\ensuremath{B\to D^{*} \ell\nu}}
\newcommand{\Btopi}{\ensuremath{B\to \pi \ell\nu}}
\newcommand{\bra}[1]{\ensuremath{\langle #1 |}}   
\newcommand{\ket}[1]{\ensuremath{| #1 \rangle}}   

\title{\large \bf 12th Workshop on the CKM Unitarity Triangle\\ Santiago de Compostela, 18-22 September 2023 \\ \vspace{0.3cm}
\LARGE 
Summary of the CKM 2023 \\Working Group on $V_{ub}$, $V_{cb}$ and semileptonic/leptonic $B$ decays including $\tau$}

\author[1]{William I. Jay\thanks{willjay@mit.edu}}
\author[2]{Raynette van Tonder\thanks{raynette.vantonder@mcgill.ca}}
\author[3]{Ryoutaro Watanabe\thanks{wryou1985@gmail.com}}
\affil[1]{Center for Theoretical Physics, Massachusetts Institute of Technology, Cambridge, MA, United States}
\affil[2]{Department of Physics, McGill University, 3600 rue University, Montr\' eal, Qu\'ebec, H3A 2T8, Canada}
\affil[3]{Institute of Particle Physics and Key Laboratory of Quark and Lepton Physics (MOE), Central China Normal University, Wuhan, Hubei 430079, China}
\maketitle


\begin{abstract}
This work summarizes recent results, both theoretical and experimental, in $B$-meson leptonic and semileptonic decays and metrology of $V_{ub}$ and $V_{cb}$, which were presented at the CKM 2023 workshop.
We place these results in context and discuss future prospects in the field.
\end{abstract} 

\maketitle
\thispagestyle{firstpage}

\section{Introduction}
Semileptonic $B$-meson decays, involving a final state with a lepton-neutrino pair, are dominated by tree-level processes in the Standard Model of particle physics (SM) and are expected to be relatively insensitive to contributions from possible new physics.
Moreover, theoretical control of semileptonic $B$ decays is greater than decays involving purely hadronic final states, due to the factorization of the leptonic and hadronic final states. 
Consequently, these relatively abundant decays offer theoretically clean avenues not only to perform precise measurements of SM parameters, but also to test lepton flavor universality involving the heavy $\tau$ lepton.
Significant experimental and theoretical progress in these areas was reported at the 2023 CKM workshop.

\section{Lepton universality}
Semileptonic $B$ decays, exhibiting clean experimental signatures and controllable theoretical uncertainties, provide an ideal toolkit to test lepton flavor universality (LFU). 
Due to the cancellation of various systematic uncertainties, measurements of ratios of decay rates can be achieved to a high precision---providing stringent tests of LFU. 
For ratios involving light leptons, LFU is well-measured by experiments and is satisfied at the level of a few percent.
However, the main driver of the current anomalies involve ratios with semitauonic decays, namely:
\begin{equation}
    R(D^{(*)}) = \frac{\mathcal{B}(B \rightarrow D^{(*)}\tau\nu_{\tau})}{\mathcal{B}(B \rightarrow D^{(*)} \ell\nu_{\ell})}\,, \qquad \ell =e\,,~\mu\,, 
\end{equation}
where $D^{(*)}$ denotes either the $D$ or $D^{*}$ meson.
The tension between the average $R(D^{(*)})$ measurements of different experiments and the corresponding SM predictions currently stands at 3.2$\sigma$~\cite{HFLAV:2022esi}. 

Apart from these results, additional measurements of various $b \rightarrow c \tau \nu $ decays and other observables can be studied to probe LFU, including $R(J/\psi)$, the $D^{*}$ longitudinal polarization, and angular asymmetries in differential decays rates.

\subsection{Tau-to-light-lepton ratios}
The talk by M.~Fedele discussed the impact of the decay $\Lambda_b \to \Lambda_c\tau\nu$ on possible new physics in $b\to c\tau\nu$ transitions~\cite{Fedele:2022iib}.
Experimental measurements from BaBar~\cite{BaBar:2012obs,BaBar:2013mob}, Belle~\cite{Belle:2015qfa,Belle:2017ilt,Belle:2019rba}, and LHCb~\cite{LHCb:2023zxo,LHCb:2023uiv,LHCb:2017vlu} for the LFU ratios of $R(D^{(*)})$ and $R(J/\psi)$ point to a deficit relative to the ratio of $R(\Lambda_c) = \mathcal{B}(\Lambda_b\to \Lambda_c \tau\bar{\nu})/\mathcal{B}(\Lambda_b\to \Lambda_c \ell\bar{\nu})$~\cite{LHCb:2022piu}.
This result is particularly interesting in light of the existence of a sum rule relating the three ratios~\cite{Blanke:2018yud,Blanke:2019qrx}.
Ref.~\cite{Fedele:2022iib} found that a common new-physics explanation of all three ratios was not possible.
Ref.~\cite{Fedele:2023ewe} subjected the hadronic form factors entering $R(D^{(*)})$ to fresh scrutiny using the dispersive matrix approach; it was not possible to simultaneously satisfy experimental constraints for $R(D^{(*)})$, the longitudinal polarization fraction of the $D^{*}$ in $B\to D^{*} \ell \nu$ with light leptons $F_L^\ell$, and the forward-backward asymmetry $A_{FB}$.

The talk by S.~Iguro discussed implications for models of new physics in light of global fits to data for $b\to c \tau \nu$ decays~\cite{Iguro:2022yzr}. The global fit points to new physics possibilities of Charged Higgs and Leptoquark scenarios. Collider signals of these scenarios~\cite{Endo:2021lhi,Blanke:2022pjy} and model constructions~\cite{Cheung:2022zsb,Iguro:2023prq}, motivated by the global fit, have also been discussed in recent years.

The Belle II experiment presented their highly anticipated first measurement of $R(D^{*})$, using $\unit[189]{fb^{-1}}$ of electron-positron collision data collected at the $\Upsilon(4S)$ resonance~\cite{Belle-II:2024ami}.
The signal $\bar{B} \rightarrow D^{*} \tau^{-} \nu_{\tau}$ and normalization $\bar{B} \rightarrow D^{*} \ell^{-} \nu_{\ell}$ channels were reconstructed using 
an experimental technique known as hadronic tagging.
Using a hierarchical multivariate algorithm, this reconstruction techniques fully reconstructs one of the two $B$ mesons produced in the $\Upsilon(4S) \rightarrow \bar{B}B$ decay through exclusive hadronic decay channels. 
The remaining charged tracks and neutral clusters in the event are used to infer the kinematics of the remaining signal $B$ meson.
As a direct consequence, this technique allows for a completeness constraint to be imposed, requiring no additional tracks in the event, which suppresses misreconstructed background. 
Furthermore, the momentum of the $B_{\textrm{tag}}$ candidate together with the precisely known initial beam-momentum, enables the missing mass squared ($M^{2}_{\textrm{miss}}$) of the event to be estimated. 
Only leptonic $\tau$ events are considered, i.e. $\tau^{-} \rightarrow e^{-}\bar{\nu}_{e}\nu_{\tau}$ and $\tau^{-} \rightarrow \mu^{-}\bar{\nu}_{\mu}\nu_{\tau}$.
The value of $R(D^{*})$ is extracted using a two-dimensional fit to $M^{2}_{\textrm{miss}}$ and the residual calorimeter energy $E_{\textrm{ECL}}$. This result is consistent with both the current world average of $R(D^{*})$ measurements and with SM predictions~\cite{HFLAV:2022esi}.
Leading sources of systematic uncertainty include the statistically limited sample size and the modeling of the $B \rightarrow D^{**} \ell \nu_{\ell}$ background decays.

The LHCb experiment performed the first simultaneous measurement~\cite{LHCb:2023zxo} of $R(D)$ and $R(D^{*})$ in hadron collision data collected in the Run-1 operating period at the LHC, which corresponds to an integrated luminosity of $\unit[3]{fb^{-1}}$ with center-of-mass energies of $\unit[7]{TeV}$ and $\unit[8]{TeV}$.
This study utilizes the purely muonic tau decay for the reconstruction of signal $\bar{B} \rightarrow D^{(*)} \tau^{-} \bar{\nu}_{\tau}$ (where $D^{(*)} = D^{0}$, $D^{*+}$, or $D^{*0}$ ) as well as normalization $\bar{B} \rightarrow D^{(*)} \mu^{-} \bar{\nu}_{\mu}$ decays that are identified using the visible final states $D^{0}\mu^{-}$ and $D^{*+}\mu^{-}$.
To reconstruct various key kinematic variables, the $B$ rest frame is approximated by assuming the $B$ boost along the $z$ axis is equal to that of the visible $D^{(*)}\mu$ candidate.
Subsequently, signal and normalization yields are then extracted from a three-dimensional binned template fit to the invariant mass of the lepton-neutrino system $q^{2}$, the muon energy in the $B$ rest frame $E_{\mu}^{*}$, and $M^{2}_{\textrm{miss}}$.
The results are consistent with the world average and agree with SM predictions with 1.9$\sigma$. LHCb also updated their measurement of $R(D^{*})$~\cite{LHCb:2023uiv} where the $\tau$ is reconstructed through hadronic modes $\tau^{+} \rightarrow 3\pi(\pi^{0})\bar{\nu}_{\tau}$, with $3\pi = \pi^{+} \pi^{-}\pi^{+}$, previously performed on the Run-1 dataset~\cite{LHCb:2017smo}. 
The new result uses data taken during 2015-2016 at $\sqrt{s} = \unit[13]{TeV}$, corresponding to $\unit[2]{fb^{-1}}$.
Despite the lower luminosity, this analysis has approximately 40\% more candidates than the previous measurement, due to improvements in the LHCb trigger system and the a higher center-of-mass energy in Run 2.
Combining this new result with the Run 1 measurement provides the most precise measurement of $R(D^{*})$ to date, which is consistent with SM predictions within 1$\sigma$.

A complementary probe of LFU, that is distinct in both statistical and systematic uncertainties to the exclusive $R(D^{(*)})$ measurements, is the tau-to-light-lepton ratio of inclusive semileptonic $B$ decays branching fractions: $R(X_{\tau/\ell}) = \mathcal{B}(B \rightarrow X\tau \nu)/\mathcal{B}(B \rightarrow X\ell \nu)$.
Belle II presented the first measurement of the inclusive ratio, studied specifically at the $\Upsilon(4S)$ resonance~\cite{Belle-II:2023aih}, using $\unit[189]{fb^{-1}}$ of the collected collision data.
Only leptonic $\tau$ events are considered, while hadronic tagging is employed to identify kinematic variables describing inclusive semileptonic $B$-meson decays and to reconstruct the hadronic $X$ system.
A data-driven approach is utilized to address the observed discrepancy between data and simulation due to mismodeling of inclusive $B \rightarrow X_{c} \ell \nu$ decays.
Both signal  $B \rightarrow X \tau \nu_{\tau}$ and normalization $B \rightarrow X \ell \nu_{\ell}$ decays are reweighted using the experimental-to-simulated yield ratio in intervals of the hadronic mass $M_{X}$ in the high-lepton-momentum control region.
This procedure enables the extraction of the signal and normalization yields with a two-dimensional fit to the lepton momentum in the rest frame of the signal $B$ meson $p^{B}_{\ell}$ and $M^{2}_{\textrm{miss}}$.
The result is found to be in agreement with an average of SM predictions~\cite{Freytsis:2015qca,Ligeti:2021six,Rahimi:2022vlv} as well as the experimental world average of $R(D^{*})$ measurements. Leading sources of systematic uncertainty include the limited size of the simulation sample, the discrepancy of branching fractions between the inclusive measurement of $\mathcal{B}(B \rightarrow X \ell \nu)$ and the sum of all exclusive final states, and the choice of different form factor parametrizations describing $B \rightarrow D^{*} \ell \nu$ decays. In preparation for this result, Belle II also studied~\cite{Belle-II:2023qyd} the inclusive ratio of branching fractions for light leptons, $R(X_{e/\mu}) = \mathcal{B}(B \rightarrow X e^{-} \nu)/\mathcal{B}(B \rightarrow X \mu^{-} \nu)$, which agreed well with unity and served as a complementary measurement to the shape test of inclusive $B \rightarrow X_{c} \ell \nu$ decays previously performed by Belle~\cite{Belle:2021idw}.

The CMS experiment presented their first LFU test in $b \rightarrow c \ell \nu$ transitions~\cite{CMS:2023vgr}, based on data collected in 2018 with a total integrated luminosity of $\unit[59.7]{fb^{-1}}$.
The analysis measured the value of
$R(J/\psi) = \mathcal{B}(B^{+}_{c} \rightarrow J/\psi \tau^{+} \nu_{\tau})/\mathcal{B}(B^{+}_{c} \rightarrow J/\psi \mu^{+} \nu_{\mu})$
by reconstructing only the muonic decay mode of the $\tau$ and $J/\psi \rightarrow \mu^{+}\mu^{-}$.
Due to the presence of neutrinos in the final state, the $B^{+}_{c}$ four-momentum can only be inferred.
By assuming that the $B^{+}_{c}$ direction of flight is aligned to that of its visible decay products, the $B^{+}_{c}$ four-momentum can be estimated as $p_{B^{+}_{c}} = m_{B^{+}_{c}}/m^{\textrm{vis}}_{3\mu} \cdot p^{\textrm{vis}}_{3\mu}$, where $m_{B^{+}_{c}}$ corresponds to the most accurate measurement of the $B^{+}_{c}$ mass, and $m^{\textrm{vis}}_{3\mu}$ and $p^{\textrm{vis}}_{3\mu}$ refer to the mass and four-momentum of the $3\mu$ candidate, respectively.
A binned maximum-likelihood fit in two variables is performed to determine the value of $R(J/\psi)$: $q^{2} = (p_{B^{+}_{c}} - p_{J/\psi})^{2}$ and $L_{xy}/\sigma_{L_{xy}}$, where $L_{xy}$ is the distance in the transverse plane between the $J/\psi$ decay vertex and the beam spot and $\sigma_{L_{xy}}$ is its uncertainty. 
The final result is found to be in agreement with the SM prediction~\cite{Harrison:2020nrv} within 0.3$\sigma$ and compatible with the measurement from LHCb~\cite{LHCb:2017vlu} within 1.3$\sigma$.
The sensitivity of the measurement is expected to improve in future iterations.

\subsection{Angular analyses}
Motivated by a reinterpretation~\cite{Bobeth:2021lya} of a Belle measurement~\cite{Belle:2018ezy} that reported a 3.9$\sigma$ tension between light-lepton modes in the angular distributions of $B^{0} \rightarrow D^{*-} \ell^{+} \nu_{\ell}$ decays, Belle II presented the first comprehensive study~\cite{Belle-II:2023svm} of angular-asymmetry observables as functions of the decay recoil $w$. These observables, called $A_{\textrm{FB}}$, $\mathcal{S}_{3}$, $\mathcal{S}_{5}$, $\mathcal{S}_{7}$ and $\mathcal{S}_{9}$, are constructed from one- or two-dimensional integrals of the $B \rightarrow D^{*} \ell \nu$ differential rate, which can be expressed in terms of $w$ and three helicity angles, $\theta_{\ell}$, $\theta_{\nu}$, and $\chi$. The differences between the angular asymmetries for electrons and muons are sensitive to interactions that violate lepton universality. A tagged analysis of $B^{0} \rightarrow D^{*-} \ell^{+} \nu_{\ell}$ is performed with $\unit[189]{fb^{-1}}$ of collision data over three $w$ ranges: the full phase-space, $w \in [1.0, 1.275]$ and $w \in [1.275, 1.5)$. The statistically limited measurements are compatible with SM expectations with a $p$-value of 0.15 over the full $w$ range.

Using the same dataset as that of the $R(D^{*})$ measurement with hadronic $\tau^{+}$ modes, LHCb showed their first measurement~\cite{LHCb:2023ssl} of the longitudinal $D^{*}$ polarization fraction in $B^0\to D^{*-}\tau^{+}\nu_{\tau}$ decays.
The value of the $F_{L}^{D^{*}}$ observable is extracted from the angular distribution of the $D^{*-} \rightarrow \bar{D}^{0}\pi^{-}$ decay, which is dependent on coefficients $a_{\theta_{D}}(q^{2})$ and $c_{\theta_{D}}(q^{2})$ that encapsulate information about hadronic effects and fundamental couplings.
These coefficients are determined from a four-dimensional fit to the $\tau^{+}$ decay time, $\cos{\theta_{D}}$ in two bins of $q^{2}$, and the output from an anti-$D^{+}_{s}$ Boosted Decision Tree (BDT).
Using the resulting coefficients, $F_{L}^{D^{*}}$ is measured in two $q^{2}$ regions, below and above $\unit[7]{GeV^{2}}$, as well as the full $q^{2}$ range. Not only are the results the most precise to date, but they are also consistent with SM predictions within 1$\sigma$.

\section{CKM metrology}
Precise determinations of the absolute value of the Cabibbo--Kobayashi--Maskawa (CKM) matrix elements provide a potent test of the SM.
A well-established strategy to determine $|V_{ub}|$ and $|V_{cb}|$ is to use measurements of semileptonic $B$-meson decays with $b \rightarrow u\ell \nu$ and $b \rightarrow c\ell \nu$ transitions.
Determinations of $|V_{ub}|$ and $|V_{cb}|$ are extracted by employing two complementary approaches: the exclusive approach focuses on the reconstruction of a specific decay mode, while the inclusive approach aims to measure the sum of all possible final states entailing the same quark-level transition.
Current world averages of $|V_{ub}|$ and $|V_{cb}|$ from exclusive and inclusive determinations exhibit disagreements of approximately 3 standard deviations between both techniques~\cite{HFLAV:2022esi}.
This disagreement has posed a longstanding puzzle.

\subsection{Combined determinations}
Combined fits incorporating updated theoretical form factors from lattice QCD (LQCD) and light-cone sum rules (LCSR) were presented by B.~Meli{\'c}~\cite{Leljak:2021vte,Leljak:2023gna} and 
C.~Bolognani~\cite{Bolognani:2023mcf}.
For the decay $B\to\pi\ell\nu$, the two theoretical frameworks play complementary roles: LQCD calculations are typically most precise in the range $19~{\rm GeV}^2 \lesssim q^2 \lesssim 25~{\rm GeV}^2$, while LCSR are most reliable for $q^2 < m_b^2 - 2m_b \Lambda \approx 15~{\rm GeV}^2$, where $\Lambda$ is a typical hadronic scale.
In Ref.~\cite{Leljak:2021vte}, a modified BCL parameterization was used as a joint model for the form factors from LQCD and LCSR, yielding $\Vub^{\rm B\to\pi\ell\nu}_{\rm LQCD + LCSR}=(3.77 \pm 0.15) \times 10^{-3}$, which agrees with the most recent determination of $\Vub^{\rm incl.}$ by the Belle Collaboration \cite{Belle:2021eni} at the $1\sigma$ level.
Although the tension between the inclusive and exclusive determination from $B\to\pi\ell\nu$ is reduced, tensions remain in other exclusive channels including $B\to\{\rho,\omega\}\ell\nu$.
This motivated the work in Ref.~\cite{Leljak:2023gna}, which analyzed all available data on exclusive semileptonic decays $b\to u\ell\bar{\nu}$ within a Bayesian framework, both in the Standard Model and in the Weak Effective Theory.
The analysis of Ref.~\cite{Leljak:2023gna} favors a beyond-the-Standard-Model interpretation over a Standard-Model interpretation to describe existing data.

Ref.~\cite{Bolognani:2023mcf} provided an updated set of $\bar{B}_s \to K$ form factors using light-cone sum rules with an on-shell kaon and combining the LQCD results at large momentum transfer using a modified BGL parameterization.
After combining with recent measurements from LHCb for the decays $B_s^0 \to \{K^-, D_s^-\}\mu^+\nu$, Ref.~\cite{Bolognani:2023mcf} determined the ratio of CKM matrix elements in regions of low and high momentum transfer, $|V_{ub}/V_{cb}|_{q^2<7~{GeV}^2} = 0.0681\pm0.0040$ and $|V_{ub}/V_{cb}|_{q^2>7~{GeV}^2} = 0.0801\pm0.0047$.

Ref.~\cite{Biswas:2022yvh} also extracted $|V_{ub}/V_{cb}|$ from a combined study of $B\to \{\pi\rho\omega\}\ell\bar{\nu}$ and $B_{(s)} \to D^{(*)}_{(s)}\ell\bar{\nu}$ decays, with many variations of the experimental data and theoretical form factors used as inputs to the fits.
Overall, Ref.~\cite{Biswas:2022yvh} reported a range of values for $|V_{ub}/V_{cb}|$, $\Vub$, $\Vcb$ in different fit scenarios.

The Belle experiment produced a new combined extraction~\cite{Belle:2023asx} of $|V_{ub}|^{\rm incl}$ and $|V_{ub}|^{\rm excl}$ using its full dataset and a hadronic tagged approach. This measurement inherits the same reconstruction strategies of previous Belle measurements of inclusive $B \rightarrow X_{u} \ell \nu$, notably suppressing the dominant $B\to X_{c} \ell \nu$ background process by utilizing a multivariate selection based on BDTs~\cite{Belle:2021eni,Belle:2021ymg}.
A two-dimensional fit is employed to extract both exclusive $B \rightarrow \pi \ell \nu$ decays and other inclusive $B \rightarrow X_{u} \ell \nu$ processes simultaneously by using the number of charged pions $N_{\pi^{\pm}}$ in the hadronic $X_{u}$ system and the squared four-momentum transfer $q^{2}$.
The analysis constrains the BCL expansion coefficients of $B \rightarrow \pi \ell \nu$ form factors to the LQCD values of Ref.~\cite{FlavourLatticeAveragingGroupFLAG:2021npn}, which combine calculations from several groups.
After including constraints based on all theoretical and available experimental knowledge of the $B \rightarrow \pi \ell \nu$ form factor shape, partial branching fractions were determined and used as input to calculate $|V_{ub}|^{\rm incl}$ and $|V_{ub}|^{\rm excl}$.
The result is compatible to the ratio of the current world averages within 1.2$\sigma$, while the calculated average of $|V_{ub}|$ is compatible with the expectation of CKM unitarity~\cite{Charles:2004jd} within 0.8$\sigma$. 

Unitarity constraints and the dispersive matrix (DM) method, which can be used in combined analyses, were discussed by L.~Vittorio~\cite{Martinelli:2022amd}.
The DM approach offers a promising avenue to constrain non-perturbative hadronic form factors while minimizing model assumptions.
Theoretical inputs for the DM method are LQCD determinations of the form factors at low recoil.
Imposing constraints from unitarity, the DM extends these results to the full kinematic range.
Consistently incorporating unitarity constraints using theoretical data with finite statistical precision presents a subtle challenge.
A new importance-sampling method, the so-called unitarity sampling procedure, was introduced in Ref.~\cite{Simula:2023ujs}.
The talk also emphasized the importance of moving toward purely theoretical determinations of form-factor shapes, using all available LQCD results in an a manner consistent with constraints from unitarity and kinematics.

\subsection{Inclusive decays}

Theoretical predictions relevant for inclusive semileptonic decays are obtained using an operator product expansion (OPE) and the heavy quark expansion (HQE).
At each order in $1/m_b$, the HQE separates the perturbatively calculable short-distance physics from non-perturbative forward matrix elements, which are typically extracted in a global simultaneous fit to experimental measurements of moments of differential distributions (e.g., charged lepton energy $E_\ell$ and hadronic invariant mass $M_X$) along with $|V_{xb}|$.

A summary of the current status of the HQE for inclusive semileptonic decays was given in the talk by K.~Vos.
A challenge of the HQE is the proliferation of non-perturbative matrix elements at higher orders. 
The challenge can be partially ameliorated using reparameterization invariance (RPI), which is the observation that physical predictions must be independent of the arbitrary choice of the vector $v$ (associated with the $B$-meson velocity) used to define the HQE~\cite{Mannel:2018mqv}.
Imposing RPI drastically reduces the number of non-perturbative inputs; for example the 13 parameters appearing at $\mathcal{O}(1/m_b^4)$ are reduced to 8 after imposing RPI. However, only the $q^2$ spectrum of $B \rightarrow X_{c} \ell \nu$ decays satisfies the RPI requirement, thus the known $E_\ell$ or $M_X$ moments can not be used in a fit to the RPI basis~\cite{Fael:2018vsp}. In this light, the Belle experiment provided the first measurement of the first four moments of the $q^2$ distribution~\cite{Belle:2021idw}, followed by a similar measurement by the Belle II experiment~\cite{Belle-II:2022evt}. These experimental inputs, in combination with the total $B \rightarrow X_{c} \ell \nu$ decay rate, are used to determine a new, complementary value of $|V_{cb}|$ together with the HQE parameters in the RPI basis~\cite{Bernlochner:2022ucr}. K. Vos also presented further work studying the contributions of $1/m^{5}_{b}$ terms in the HQE to estimate the impact of theoretical uncertainties for future inclusive determinations of $|V_{cb}|$~\cite{Mannel:2023yqf}.

Inclusive $B\to X_u$ decays present a challenge due to large backgrounds from $B\to X_c$.
Experiments remove these backgrounds with kinematic cuts, which cannot be described in a local OPE. Consequently, theoretical calculations of distributions in $B \rightarrow X_{u} \ell \nu$ decays introduce non-local OPEs, leading to non-perturbative distribution functions, the so-called shape functions~\cite{Bigi:1993ex,Neubert:1993um,Neubert:1993ch}.
The moments of these shape functions are related to the non-perturbative parameters of the HQE. 
Preliminary results of ongoing work by K.~Vos and collaborators improve upon the existing calculations by not only including known $\mathcal{O}(\alpha_s^{2})$ corrections and constraining the higher moments of the shape functions, but also using the kinematic mass scheme for the bottom quark mass.
This approach is advantageous, since the kinematic scheme exhibits better perturbative properties than the pole-mass scheme.

Further frontiers of the HQE were discussed in other talks throughout the CKM workshop. D.~Moreno reported on recent calculations of the next-to-leading-order QCD corrections to the spectrum and decay rate for inclusive $B\to X_c \tau \nu$ up to $\mathcal{O}(1/m_b^3)$.
In the heavy-quark expansion, corrections of $\mathcal{O}(1/m_b^3)$ provide $\approx 10\%$ shift to the leading term.
Contributions of $\mathcal{O}(\alpha_s/m_b^2)$ and $\mathcal{O}(\alpha_s/m_b^3)$ were calculated analytically in Ref.~\cite{Moreno:2022goo,Mannel:2021zzr} and furnish further $\approx 1\%$ corrections, which are expected to provide a small but visible impact on \Vcb.

M.~Fael described calculations of new-physics contributions to moments of inclusive $b\to c$ decays~\cite{Fael:2022wfc}. 
The goal, facilitated by new open-source software packages \href{https://gitlab.com/vcb-inclusive/npinb2xclv}{\texttt{npinb2xclv}} and \href{https://gitlab.com/vcb-inclusive/kolya}{\texttt{kolya}}, is comprehensive model-independent analysis of all possible types of new-physics effects in $B\to X_c \ell\nu$.
Doing so requires considering a series expansion in three parameters: $\Lambda_{\rm QCD}/m_b$, $\alpha_s$, and $(v/\Lambda_{NP})^2$.
Here $\Lambda_{\rm QCD}$ and $\Lambda_{\rm NP}$ are the scales of QCD and new physics, respectively, and $v=246~{\rm GeV}$ is the Higgs vacuum expectation value.
Results were presented for charged-lepton energy moments (with lower cut on lepton energy $E_\ell$), hadronic invariant mass moments (with lower cut on lepton energy $E_\ell$), and leptonic invariant mass moments (with lower cut on momentum transfer $q^2$).
Regarding the search for new physics, the electron energy spectrum and forward-backward asymmetries are viewed as promising observables~\cite{Turczyk:2016kjf,Herren:2022spb}.
Observables like moments of the hadronic invariant mass $M_X$ or the momentum transfer $q^2$ are more sensitive to higher-order corrections in the HQE, making it more difficult to disentangle possible effects from new physics.~\cite{Fael:2022wfc}

Motivated the long-standing tension between inclusive and exclusive determinations of $\Vcb$ and $\Vub$, work is underway to compute inclusive semileptonic decays (e.g., $B_s\to X_s \ell \nu$) using lattice QCD.
The challenge is both algorithmic and conceptual.
The basic hadronic input for inclusive $B$-meson decay is a set of Euclidean four-point correlation functions of the form $\bra{B}J_\mu^\dagger(t, \bm{q}) J_\nu(0, -\bm{q}) \ket{B}$, where $J_\mu(t, \bm{q})$ is the weak current in the time-momentum representation.
Computing such four-point functions is numerically more intensive than the two- and three-point functions required for leptonic and exclusive semileptonic decays.
The fundamental theoretical challenge is relating the (Euclidean) time dependence to the energy transfer in the decay, a problem which amounts to computing a numerically delicate inverse Laplace transform.
Ref.~\cite{Hansen:2017mnd} showed that this problem could be regulated by a suitable smearing, which Ref.~\cite{Gambino:2020crt} argued could be provided by the phase-space integral for the total rate.
R.~Kellermann gave an updated on recent progress in lattice QCD calculations of $B$ and $D$ mesons~\cite{Gambino:2022dvu,Gambino:2022xww,Smecca:2022bcq,Barone:2022gkn,Kellermann:2022mms,Kellermann:2023yec,Barone:2023iat,Barone:2023tbl}.
Development of inclusive calculations are proceeding quickly within the lattice community, but several years will likely be required before the same level of precision and systematic control as seen in exclusive semileptonic decays are achieved.
A particular focus is quantifying finite-lattice-size effects~\cite{Kellermann:2023yec}, which are expected to appear with a power-law dependence on the volume ($\propto1/L^N$ for some $N$) as opposed to the exponential suppression ($\propto e^{-M_\pi L}$) appearing in matrix elements between single-hadron states.

\subsection{Exclusive decays}

A series of exciting exclusive semileptonic measurements has been performed and presented since the previous CKM conference. Both the Belle and Belle II Collaborations published new differential spectra for the $B \rightarrow D^{*} \ell \nu$ decay~\cite{Belle:2023bwv,Belle-II:2023okj}, while also studying various quantities sensitive to lepton flavor universality violation, such as the longitudinal $D^{*}$ polarization, $F_{L}^{D^{*}}$, and the charged-lepton forward-backward asymmetry, $A_{\textrm{FB}}$. These variables were found to be consistent with their respective SM predictions.

With $\unit[189]{fb^{-1}}$ of collected collision data, Belle II studied $B^{0} \rightarrow D^{*-} \ell^{+} \nu$ decays by making use of an untagged approach, i.e only the signal $B^{0}$ ($B_{\textrm{sig}}$) is reconstructed. The direction of the signal $B$ is determined by combining angular information from the $Y = D^{*}\ell$ system with information from the remaining tracks and neutral clusters in the event not associated with the $D^{*}\ell$. With the direction of $B_{\textrm{sig}}$, distributions of the recoil parameter $w$ and three the helicity angles $\theta_{\ell}$, $\theta_{\nu}$ and $\chi$ can be determined. Partial decay rates are then measured as functions of the four differential kinematic variables. A value of $|V_{cb}|$ is determined by making use of a nested hypothesis test~\cite{Bernlochner:2019ldg} to determine the order at which to truncate the expansion of the BGL form factors. Their result is in good agreement with the world average of the exclusive approach and the inclusive determinations of Refs.~\cite{Bordone:2021oof,Bernlochner:2022ucr}.
Additionally, the analysis investigated the effect of including non-zero recoil information from Fermilab--MILC~\cite{FermilabLattice:2021cdg} for two different scenarios: considering only $h_{A_{1}}$ yielded a slight decrease in the calculated value of $|V_{cb}|$, whereas the inclusion of all form factors showed a tension with LQCD predictions in Ref.~\cite{FermilabLattice:2021cdg}.

Belle also analyzed the differential shapes of $B \rightarrow D^{*} \ell \nu$ decays with its complete dataset, while using hadronic tagging to reconstruct fully the companion $B$.
In contrast to the previous tagged analyses of this decay, the $B^{+}$ mode is also considered, which significantly increases the available statistics in the low-hadronic-recoil region.
This analysis relies on external branching fractions provided by HFLAV to extract $|V_{cb}|$ by fitting the measured kinematic differential shapes.
The number of floating BGL parameters is also determined by making use of a nested hypothesis test.
The resulting value of $|V_{cb}|$ is consistent regardless of the chosen form factor parametrization (BGL~\cite{Boyd:1995cf,Boyd:1995sq,Boyd:1997kz} or CLN~\cite{Caprini:1997mu}), agrees well with the Belle II measurement, and is compatible with inclusive determinations.
Additionally, the analysis is extended to perform the first measurement of the complete set of normalized angular coefficients~\cite{Belle:2023xgj}, which encode the full angular information of the $B \rightarrow D^{*} \ell \nu$ decay and provide a more comprehensive set of observables than the one-dimensional partial rates.
Under the assumption that physics beyond the SM does not contribute to these modes, a combined fit with LQCD calculations at nonzero-recoil (discussed more below) by Fermilab--MILC~\cite{FermilabLattice:2021cdg}, HPQCD~\cite{Harrison:2023dzh} and JLQCD~\cite{Aoki:2023qpa} for both CLN and BGL form factor parametrizations is performed to extract $|V_{cb}|$ from the measured angular coefficients.
Consistent values of $|V_{cb}|$ are determined for both parametrizations, while good agreement for different LQCD inputs is also observed. 

For many years, LQCD calculations of the hadronic form factors for \BtoDstar~were limited to zero recoil.
The situation changed in 2021, when the Fermilab Lattice and MILC collaborations computed the form factors in the whole recoil range Ref.~\cite{FermilabLattice:2021cdg}. 
Progress has continued since the last CKM workshop, with two additional calculations spanning the full kinematic range by HPQCD~\cite{Harrison:2023dzh} and JLQCD~\cite{Aoki:2023qpa}.
As mentioned in the previous paragraph, these LQCD results are already being utilized in extractions of \Vcb.
Quantities involving heavy quarks have historically presented a difficult challenge for LQCD, primarily related to controlling discretization effects with a physical scale at or beyond the lattice cutoff ($m_b a \gtrsim 1$).
The three recent calculations used complementary methods, which are summarized in Table~\ref{tab:B2Dstar_comparison}.
In Ref.~\cite{FermilabLattice:2021cdg}, the charm and bottom quarks are treated using HQET, allowing for $m_h/m_b\approx 1$.
In Refs.~\cite{Harrison:2023dzh,Aoki:2023qpa}, the charm and bottom quarks are treated using relativistic light-quark actions; for the lattice spacings used in these calculations, this treatment required $m_h/m_b < 1$.

\begin{figure}[h]
    \centering
    \includegraphics[width=0.75\textwidth]{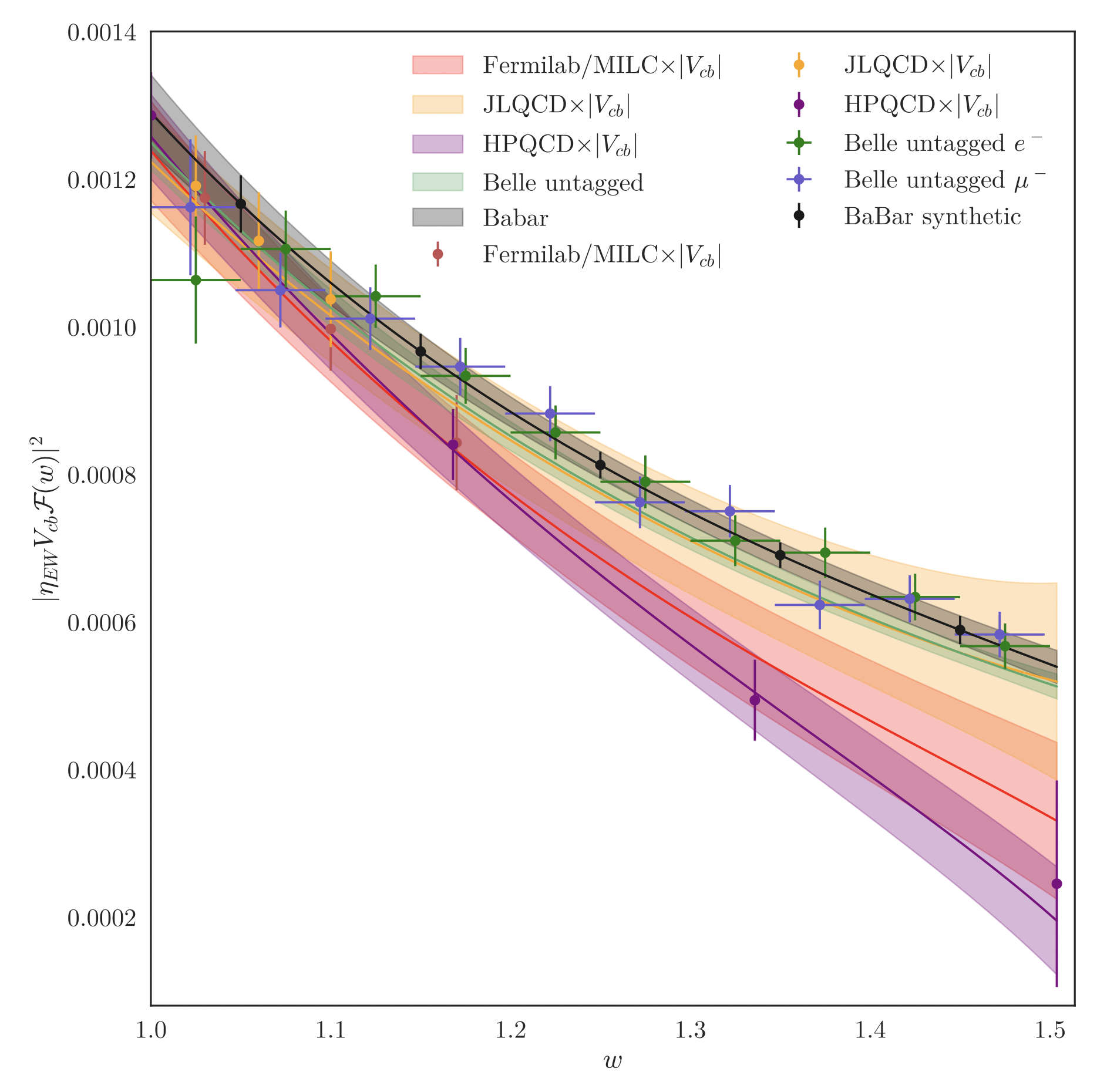}
    \caption{Comparison between lattice QCD calculations and experimental measurements for the combination of hadronic form factors that appear in the differential decay rate for the decay $B\to D^{*}\ell\nu$.
    The figure is from the talk by A.~Vaquero.
    \label{fig:vaquero_b2dstar_comparison}}
\end{figure}

The level of agreement between the different calculations was addressed in the talk by A.~Vaquero, which indicated that the three different calculations could be combined reasonably in a joint fit.
\Cref{fig:vaquero_b2dstar_comparison} shows one such comparison for the combination of form factors that appears in the differential decay rate for \BtoDstar.
The red, yellow, and purple curves in \cref{fig:vaquero_b2dstar_comparison} are from lattice QCD calculations, while the black and green curves are from BGL fits to experimental data.
As similar situation is present in comparisons of the differential decay rate $d\Gamma/dw$ between experimental measurements and LQCD, where the predictions from Refs~\cite{FermilabLattice:2021cdg,Harrison:2020nrv} differ somewhat from the experimentally measured shapes.

\begin{table}[t]
    \centering
    \caption{Comparison of complementary methods used in recent lattice QCD calculations of hadronic form factors for $B\to D^{*} \ell \nu$ spanning the kinematic range.
    For the quark actions, asqtad is the $a^2$ tadpole-improved action, HISQ is the highly-improved staggered quark action, and MDWF is the M{\"o}bius domain-wall fermion action.\\
    \label{tab:B2Dstar_comparison}
    }
    \begin{tabular}{c|ccc}
    \hline\hline
                                & Fermilab-MILC~\cite{FermilabLattice:2021cdg}                         & HPQCD~\cite{Harrison:2023dzh}                        & JLQCD~\cite{Aoki:2023qpa}\\
    \hline
    Sea-quark action            & asqtad                      & HISQ             & MDWF \\
    Dynamical flavors $N_f$     & $2+1$                    & $2+1+1$ & $2+1$\\
    Heavy-quark treatment       & Wilson-clover                         & HISQ                         & MDWF \\
    Lattice spacing $\approx a$ [fm] & $0.044 - 0.15$               & $0.044 - 0.09$      & $0.044 - 0.08$\\
    Pion mass $M_\pi~[{\rm MeV}]$ & $\gtrsim 180$               & $\gtrsim 135$      & $\gtrsim 230$ \\ 
    Heavy-quark mass $m_h/m_b$          & $\approx 1$                         & $\lesssim 0.9$            &  $\lesssim 0.75$ \\
    \hline\hline
    \end{tabular}
\end{table}

As a first step to understanding $B \rightarrow D^{**} \ell \nu$ decays, which constitute a significant and mostly unmeasured background for LFU tests or studies of rare processes like $B \rightarrow K \nu \nu$~\cite{Belle-II:2023esi}, Belle reported ratios of branching fractions for $B \rightarrow \bar{D}^{(*)} \pi \ell^{+} \nu_{\ell}$ and $B \rightarrow \bar{D}^{(*)} \pi^{+}\pi^{-}\ell^{+} \nu_{\ell}$ relative to $B \rightarrow \bar{D}^{*} \ell^{+} \nu_{\ell}$ decays~\cite{Belle:2022yzd}, using the full Belle dataset.
Hadronic tagging is implemented to reconstruct signal decays of interest by combining one $D^{(*)}$ candidate, one lepton candidate, and one charged pion candidate.
In total, 16 final states are considered. Signal and normalization candidates are determined with an unbinned extended maximum likelihood fit to the $U = E_{\textrm{miss}} - p_{\textrm{miss}}$ distribution.
The results are the most precise determinations of these branching fraction ratios to date (except for $B \rightarrow \bar{D}^{*-} \pi^{+}\pi^{-}\ell^{+} \nu_{\ell}$).
Additionally, the $D^{**}$ composition is studied by performing weighted unbinned maximum likelihood fits to the invariant-mass distributions $m(D\pi)$, $m(D^{*}\pi)$ and $m(D\pi \pi)$.
Several exclusive $B \rightarrow D^{**} \ell \nu$ branching fractions are extracted, including the first observation of $B \rightarrow D_{1} \ell^{+}\nu_{\ell}$ with $D_{1} \rightarrow D \pi^{+} \pi^{-}$. 
While the results for the decays via the narrow $D_{1}$, $D_{2}^{*}$ resonances are compatible with world averages, the value for $\mathcal{B}(B^{+} \rightarrow \bar{D}_{0}^{*}\ell^{+}\nu_{\ell}) \times \mathcal{B}(\bar{D}_{0}^{*0} \rightarrow D^{-}\pi^{+})$ is found to be significantly smaller than previous measurements. A similar trend is observed for the wider $D_{1}^{\prime}$ resonance where the branching fractions are measured 34\% (50\%) lower than the world average in the $B^{0}$ ($B^{+}$) mode.

These results were directly applicable to a new model-independent parameterization of $B\to D\pi \ell \nu$ decays that was presented by F.~Herren~\cite{Gustafson:2023lrz}.
The new parameterization generalizes the formalism developed in Refs.~\cite{Boyd:1995cf,Boyd:1995sq,Boyd:1997kz} to semileptonic decays with multi-hadron final states.
The resulting form factors can be determined in a data-driven manner, with robust, systematically-improvable uncertainties.
The $D\pi$ invariant-mass spectrum in $B\to D\pi\ell\nu$, recently measured by Belle~\cite{Belle:2022yzd}, is well described using a fit to the new parameterization.
The fit also yields new for predictions for related quantities, for instance the branching fraction for $D$-wave $B\to D_2^{*}(\to D\pi)\ell\nu$.
In addition, for the first time, input from $D\pi$ scattering is used to describe the $S$-wave, while a coupled channel treatment of the $S$-wave allows for predictions of the branching ratios of $B \rightarrow D^{(*)} \eta \ell \nu$ decays to be inferred.
The branching ratios are found to be of the order $10^{-5}$ and therefore can not account for the gap between the inclusive $B \rightarrow X \ell \nu$ branching fraction and the sum over known exclusive states.

Form factors relevant for exclusive semileptonic decays can also be treated using heavy-quark effective theory.
M.~Prim reported on recent work using the residual chiral expansion (RCE) conjecture~\cite{Bernlochner:2022ywh}. 
Working within heavy-quark effective theory, the RCE conjecture starts from the observation that matrix elements involving many insertions of the transverse residual momentum $\slashed{D}_\perp$ are often small, promoting the observation to a supplemental power-counting scheme: $\slashed{D}_\perp \sim \theta$.
Counting insertions of $\slashed{D}_\perp$ furnishes an additional classification of terms besides the $1/m_Q$ expansion.
Using the RCE power counting to truncate the expansion reduces the number of non-perturbative parameters arising at a given order.
M.~Prim reported updates for the form factors for $B\to D^{(*)}\ell\nu$ in the heavy-quark effective theory using the RCE conjecture.
It was further discussed how $\Lambda_b \to \Lambda_c \ell \nu$, which only includes two subleading Isgur-Wise functions at $\mathcal{O}(1/m_c^2)$, provides an appealing system in which to test the consistency of the RCE conjecture~\cite{Bernlochner:2023jkp}.

The talk by B.~Colquhoun discussed JLQCD's recent calculation of form factors for \Btopi~\cite{Colquhoun:2022atw}.
Similar to JLQCD recent calculation of \BtoDstar, Ref.~\cite{Colquhoun:2022atw} used a relativistic light-quark action (M{\"o}bius domain-wall fermions) for all quarks, including the heavy quark approaching the bottom quark.
This calculation used valence heavy quarks with $m_h/m_c \leq 2.44$ and an extrapolation to the physical bottom mass.
When combined with experimental data the decay rate, Ref.~\cite{Colquhoun:2022atw} reported $\Vub^{\Btopi}=3.93(41)\times 10^{-3}$.
This value is consistent at one standard deviation with extractions using previous lattice QCD calculations, as well as with both the inclusive and exclusive determinations reported by HFLAV~\cite{HFLAV:2022esi}.

Much focus within the lattice QCD community has been on weak decays involving zero or one hadron in the final state (e.g., $B_s^0 \to \mu^+\mu^-$, $B_s\to D_s \ell\nu$).
Recent theoretical and algorithmic work is laying the foundation for lattice QCD calculations of exclusive processes with multiple final-state hadrons.
Compared to decays with single-hadron states, the essential technical complication is the non-trivial relationship between multi-hadron states in a finite volume and those in an infinite volume~\cite{Lellouch:2000pv}.
Relevant lattice QCD calculations have been carried already for $K\to\pi\pi$~\cite{Blum:2011ng,RBC:2015gro,Ishizuka:2018qbn,RBC:2020kdj,RBC:2021acc}.
Weak decays of bottom mesons have many kinematically accessible final states, which make numerical application of the relevant theoretical formalism more challenging.
The talk by L.~Leskovec reviewed recent progress toward lattice calculations of semileptonic decays involving resonances, e.g., $B\to K^{*}(\to K\pi)\ell^+ \ell^-$ and $B \to \rho(\to\pi\pi)\ell\nu$~\cite{Briceno:2021xlc,Leskovec:2022ubd,Leskovec:2024pzb}.
In particular, Ref.~\cite{Leskovec:2022ubd} showed preliminary results for the first unquenched LQCD calculation of $B \to \rho(\to\pi\pi)\ell\nu$.

LQCD calculations also enter in the Standard-Model predictions for rare loop-mediated processes $b\to s$ and $b\to d$, which occur in the decays $B\to K\ell^+\ell^-$ and $B\to\pi\ell^+\ell^-$.
Compared to the tree-level charged-current processes, the SM prediction for these decays involves the tensor form factor $f_T$ in addition to the vector and scalar form factors ($f_+$ and $f_0$, respectively).
In 2023, HPQCD published a calculation of the scalar, vector, and tensor form factors for the decays $B/D\to K\ell^+\ell^-$ across the full kinematic range~\cite{Parrott:2022rgu}.
Their calculation was the first to a relativistic light-quark action for the bottom quark in $B\to K \ell^+\ell^-$.
They reported a substantial improvement in precision (roughly a factor of three) compared to calculations from 7--10 years ago.
An important technical aspect of the calculation is precise non-perturbative renormalization of the tensor current, which is facilitated by the use of the HISQ action for all quarks in the calculation~\cite{Hatton:2020vzp}.
Standard Model predictions for 
$B\to K\ell^+\ell^-$, $B\to K \ell_1^- \ell_2^+$ and $B\to K\nu \bar{\nu}$ were presented in~\cite{Parrott:2022zte} using the form factors calculated in Ref.~\cite{Parrott:2022rgu}.
Notable tensions were observed with experimental results for the branching fraction of $B\to K \ell^+\ell^-$.
The tensions can be reduced through shifts in the Wilson coefficients $C_9$ and $C_{10}$ in the effective weak Hamiltonian.

Looking forward, M.~Rotondo provided a comprehensive overview of future prospects at the LHCb experiment.
Several existing semileptonic results use the LHCb Run-1 or Run-2 dataset, but efforts are underway to update measurements using either Run-2 data or the full LHCb dataset from Run 1 and Run 2.
Furthermore, the expected increase of statistics and improved detector performance foreseen in Run 3 and Run 4 with Upgrade 1, coupled with improvements to fast simulation techniques~\cite{Muller:2018vny}, will significantly improve statistical and systematic uncertainties in measurements of $R(D^{(*)})$.
While precision studies of large background modes will also reduce the systematic uncertainty associated with the background processes, external inputs provided by Belle II and BES III will be crucial.
In addition to measurements of $R(D^{*})$ shown at this conference, several exclusive analyses are also underway: $R(D^{0})$, $R(D^{+})$, $R(D^{+}_{s})$ and $R(\Lambda_{c}^{+})$.
Decays involving $b \rightarrow u \ell \nu$ transitions are being investigated, such as $\Lambda_{b}^{0} \rightarrow p \tau^{-}\bar{\nu}_{\tau}$ and $B^{+} \rightarrow p\bar{p}\tau^{+}\nu_{\tau}$. Angular analyses and measurements of differential distributions provide complementary information to that of ratios of branching fractions, which allow for different New Physics models to be distinguished and detected, even in the case where $R(D^{(*)})$ becomes compatible with SM predictions. A measurement of the longitudinal $D^{*}$ polarization  in $B^{0} \rightarrow D^{*-} \tau^{+} \nu_{\tau}$ decays is currently underway as well as other angular analyses that measure additional observables, proposed in several methods e.g, Ref.~\cite{Hill:2019zja}. Additionally, precise measurements of exclusive $B_{s} \rightarrow H_{cs}^{**} \mu \nu$ decays would allow for the determination of the hadronic moments in inclusive $B^{0}_{s} \rightarrow X_{cs} \mu \nu$ decays using a Sum-of-Exclusive Modes approach~\cite{DeCian:2023ezb}.

\section{Outlook}

The persistent anomalies in decays of $b$-quarks, with their hint of possible new physics, are a major driver for the world-wide efforts in flavor physics.
Substantial work and progress in the this area, both theoretical and experimental, was reported at the 2023 CKM workshop and was summarized in the present work.
Despite this progress, tensions remain unresolved in inclusive versus exclusive determinations of \Vcb and \Vub, branching fractions and angular observables in neutral-current $b\to s\mu^+\mu^-$ decays, and the LFU ratios $R(D)$ and $R(D^{*})$.
This situation underscores the urgent need for new, precise experimental measurements and accompanying theoretical calculations with commensurate improvement in systematic control.\\


{\bf Acknowledgements.} We thank the speakers and organizers of CKM 2023 for their excellent work. RvT acknowledges support by the Natural Sciences and Engineering Research Council of Canada.
WJ is supported in part by the U.S. Department of Energy, Office of Science under grant Contract Numbers DE-SC0021006 and DE-SC0011090.
This work is supported by the National Natural Science Foundation of China under Grant Nos.~12135006 and 12075097, as well as by the Fundamental Research Funds for the Central Universities under Grant No.~CCNU22LJ004.

\bibliographystyle{jhep}
\bibliography{refs.bib}

\end{document}